\documentclass[sigconf,screen, nonacm, authorversion]{acmart}

\usepackage{booktabs} 
\usepackage{array} 
\usepackage{longtable} 
\usepackage{colortbl}
\usepackage{booktabs}
\usepackage{soul}

\AtBeginDocument{%
  }

\copyrightyear{2025}
\acmYear{2025}
\setcopyright{cc}
\setcctype{by}
\acmConference[FSE Companion '25]{33rd ACM International Conference on the Foundations of Software Engineering}{June 23--28, 2025}{Trondheim, Norway}
\acmBooktitle{33rd ACM International Conference on the Foundations of Software Engineering (FSE Companion '25), June 23--28, 2025, Trondheim, Norway}\acmDOI{10.1145/3696630.3731666}
\acmISBN{979-8-4007-1276-0/2025/06}


\begin{document}

\title{Get on the Train or be Left on the Station: \\ Using LLMs for Software Engineering Research}

\author{Bianca Trinkenreich}
\email{bianca.trinkenreich@colostate.edu}
\affiliation{%
  \institution{Colorado State University}
  \city{Fort Collins}
  \country{USA}
}

\author{Fabio Calefato}
\email{fabio.calefato@uniba.it}
\affiliation{%
  \institution{University of Bari}
  \city{Bari}
  \country{Italy}
}

\author{Geir Hanssen}
\email{ghanssen@sintef.no}
\affiliation{%
  \institution{SINTEF}
  \city{Trondheim}
  \country{Norway}
}

\author{Kelly Blincoe}
\email{k.blincoe@auckland.ac.nz}
\affiliation{%
  \institution{University of Auckland}
  \city{Auckland}
  \country{New Zealand}
}

\author{Marcos Kalinowski}
\email{kalinowski@inf.puc-rio.br}
\affiliation{%
  \institution{PUC-Rio}
  \city{Rio de Janeiro}
  \country{Brazil}
}

\author{Mauro Pezzè}
\email{mauro.pezze@usi.ch}
\affiliation{%
  \institution{USI Università della Svizzera Italiana}
  \city{Lugano}
  \country{Italy}
}

\author{Paolo Tell}
\email{pate@itu.dk}
\affiliation{%
  \institution{IT University of Copenhagen}
  \city{Copenhagen}
  \country{Denmark}
}

\author{Margaret-Anne Storey}
\email{mstorey@uvic.ca}
\affiliation{%
  \institution{University of Victoria}
  \city{Victoria}
  \country{Canada}
}

\renewcommand{\shortauthors}{Trinkenreich et al.}

\begin{abstract}

 The adoption of Large Language Models (LLMs) is not only transforming software engineering (SE) practice but is also poised to fundamentally disrupt how research is conducted in the field. 
 While perspectives on this transformation range from viewing LLMs as mere productivity tools to considering them revolutionary forces, we argue that the SE research community must proactively engage with and shape the integration of LLMs into research practices, emphasizing human agency in this transformation. 
 As LLMs rapidly become integral to SE research—both as tools that support investigations and as subjects of study—a human-centric perspective is essential. Ensuring human oversight and interpretability is necessary for upholding scientific rigor, fostering ethical responsibility, and driving  advancements in the field.
 Drawing from discussions at the 2nd Copenhagen Symposium on Human-Centered AI in SE, this position paper employs  McLuhan's Tetrad of Media Laws to analyze the impact of LLMs on SE research. 
 Through this theoretical lens, we examine how LLMs enhance research capabilities through accelerated ideation and automated processes, make some traditional research practices obsolete, retrieve valuable aspects of historical research approaches, and risk reversal effects when taken to extremes. 
 Our analysis reveals opportunities for innovation and potential pitfalls that require careful consideration. 
 We conclude with a call to action for the SE research community to proactively harness the benefits of LLMs while developing frameworks and guidelines to mitigate their risks, to ensure continued rigor and impact of research in an AI-augmented future.
\vspace{-3mm}
\end{abstract}


\maketitle
\vspace{-3mm}
\section{Introduction}

Integrating Large Language Models (LLMs) into Software Engineering (SE) research reflects a broader transformation across scientific disciplines. Generative AI technologies are fundamentally changing how research is conducted, from accelerating hypothesis generation to enhancing data analysis and interpretation~\cite{wang_scientific_2023}. This transformation is particularly relevant for SE research, where LLMs are becoming integral both as subjects of our investigations and as tools we use to conduct research. These models have demonstrated their potential to revolutionize research in our field by supporting various tasks, such as enhancing brainstorming processes~\cite{storey2024disruptive}, generating representative data~\cite{steinmacher2024can,gerosa2024can}, aiding in data analysis and qualitative research~\cite{BarrosAGKKNB25}, and automating repetitive or tedious tasks. 

As SE researchers increasingly incorporate LLMs into their workflows, it becomes crucial to maintain a human-centric perspective, particularly when studying human aspects of SE~\cite{russo2024manifesto}. 
The transformative potential of LLMs extends beyond mere automation as these tools can augment our ability to understand developer experiences, team dynamics, and socio-technical interactions in software development. However, this potential must be balanced against the need to preserve human agency and ensure that our research methods remain rigorous, transparent, and ethically sound. This is particularly important as we study how software developers adapt to and integrate LLMs into their work practices, requiring us to critically examine our own use of these tools in researching such phenomena. Therefore, understanding the broad impact of LLMs requires a comprehensive framework that evaluates their benefits and potential unintended consequences.

Marshall McLuhan’s Tetrad of Media Effects~\cite{mcluhan1977laws} provides a compelling lens through which to examine the different ways a new medium may have on augmenting human abilities across several dimensions. The Tetrad prompts us to critically assess technologies by addressing four key questions: What does the technology enhance? What does it make obsolete? What does it retrieve? And what does it reverse into when taken to extremes? As McLuhan observed, recognizing what a technology retrieves can be challenging, as it demands a deep historical understanding of its predecessors \cite{mcluhan2017medium}. Speculating what a technology may reverse into can also be difficult, especially when a technology is novel and disruptive. 

Inspired by previous applications of McLuhan's Tetrad to disruptive technologies~\cite{storey2024disruptive}, we applied the framework to speculate about the effects when we use LLMs in SE research. This speculation was conducted collaboratively by a team of 10 researchers during the 2nd Symposium on Human-Centered AI in SE. We note that other researchers may have different ideas about the effects LLMs will have on SE research and that our speculation would also change as LLMs evolve.

This position paper explores the multifaceted role of LLMs in SE research and discusses potential risks, including biases, creativity echo chambers, and a decline in essential research skills.
By applying the Tetrad across the Research Pipeline Stages, this paper examines LLMs from the perspectives of enhancement, obsolescence, retrieval, and reversal for each stage of the pipeline, offering a structured and reflective analysis of their impact on SE research. 

\vspace{-2mm}
\section{McLuhan's Tetrad for SE Research}

We discuss the impact of LLMs on SE research through the lens of Marshall McLuhan’s Tetrad of Media Laws (Enhance, Obsolesce, Retrieve, and Reverse) \cite{mcluhan1977laws} for each stage of a generic research pipeline, summarized in Table~\ref{tab:tetrad_llms_se}. The structure of Table ~\ref{tab:tetrad_llms_se}, including the research pipeline phases, was generated using a GPT associated with the Disruptive Playbook for studying the effects of technologies on SE~\cite{storey2024disruptive}, then refined and filled in by the authors.
In this section, we expand on two stages of the research pipeline pertinent to Human-Centric aspects of SE research: \textsc{Research Goals and Questions Formulation} and \textsc{Analysis and Interpretation}. 

\begin{table*}
\vspace{-3mm}
\footnotesize
    \centering
    \renewcommand{\arraystretch}{1.3} 
    \setlength{\tabcolsep}{3pt} 
    \begin{tabular}{p{2cm} p{3.5cm} p{3.5cm} p{3.5cm} p{3.5cm}} 
    \toprule
    \rowcolor[gray]{0.9} 
        \multicolumn{1}{c}{\textbf{Research}}
        & \multicolumn{1}{c}{\textbf{Enhance}}
        & \multicolumn{1}{c}{\textbf{Obsolesce}}
        & \multicolumn{1}{c}{\textbf{Retrieve}}
        & \multicolumn{1}{c}{\textbf{Reverse}}
    \\
    \rowcolor[gray]{0.9} 
        \multicolumn{1}{c}{\textbf{Pipeline Stage}}
        & \multicolumn{1}{c}{\scriptsize{What does it amplify?}}
        & \multicolumn{1}{c}{\scriptsize{What does it push aside?}}
        & \multicolumn{1}{c}{\scriptsize{What does it bring back?}}
        & \multicolumn{1}{c}{\scriptsize{What happens when pushed to extremes?}}
    \\
    \\[-5pt] 
        \textbf{Research Goals \newline \& Questions \newline Formulation (*)}
        & Rapid idea generation (*), auto-suggested hypotheses, literature summarization automation
        & Manual literature review (*), brainstorming without AI assistance
        & ``Coffee house research" (*), switch trending research topics (*), sketchbook of ideas 
        & Creativity echo chamber (*), homogenized research questions, potential loss of novelty from AI ideas
    \\ \midrule[.1pt]
        \textbf{Experimental \newline Design \& \newline Methodology}
        & Automated experiment setup, code synthesis for study prototypes, reproducibility improvements
        & Tedious manual setup, reliance on domain experts for experiment structuring
        & Human "intractable" models of research field, modular and reusable experimental designs
        & Over-reliance on AI-generated methodologies may lead to reduced critical evaluation
    \\ \midrule[.1pt]
        \textbf{Data \newline Collection}
        & Faster extraction from repositories (GitHub, Stack Overflow), automated data cleaning
        & Human-driven data curation, traditional data wrangling techniques
        & Historical datasets revisited for new insights
        & Bias amplification in datasets, lack of transparency in synthetic data creation
    \\ \midrule[.1pt]
        \textbf{Data Processing}
        & Improved statistical modeling via AI, anomaly identification
        & Pushing aside the risk of human errors
        & Large-scale or longitudinal ethnographic studies
        & Errors and biases that humans cannot easily detect
    \\ \midrule[.1pt]
        \textbf{Analysis \& \newline Interpretation (*)}
        & Qualitative, quantitative, and mixed-methods analysis (*), diverse viewpoints (*)
        & Manual coding (qualitative and quantitative) (*), manual selection and execution of statistical techniques (*)
        & Finding related theories in other domains (*), holistic and interdisciplinary analysis (*)
        & AI hallucinations (*), loss of human's role in theories' construction,  misleading interpretations if results are blindly trusted
    \\ \midrule[.1pt]
        \textbf{Writing \& \newline Dissemination}
        & Automated paper drafting, AI-assisted summaries, multilingual dissemination
        & Manual academic writing, sole reliance on human synthesis
        & Collaborative, rapid prototyping of research papers
        & Proliferation of low-quality or AI-generated papers, diminishing originality and rigor
    \\
    \midrule[.1pt]\bottomrule[.1pt] 
        \textbf{Cross-cutting \newline Impacts (*)}
        & Research speed and creativity (*)
        & Manual/tedious research tasks (*)
        & Impactful research (*)
        & Lower skills of researchers (*)
    \\ \bottomrule 
    \end{tabular}
    \caption{Applying McLuhan's Tetrad to LLMs across the Software Engineering Research Pipeline. (*) are discussed in this paper.}
    
    \label{tab:tetrad_llms_se}
\vspace{-7mm}
\end{table*}

\vspace{-2mm}
\subsection{Research Goals and Questions Formulation}



The formulation of research goals and questions represents a foundational yet often rushed phase of the research process, where initial decisions are frequently determined prematurely without sufficient exploration of the problem space~\cite{storey2024disruptive}. While traditionally reliant on manual literature review and brainstorming, this phase is being transformed by the capabilities of LLMs.

\textit{\textbf{Enhance:}} LLMs can amplify \textsc{rapid idea generation} and accelerate the initial phase of research, as researchers can quickly explore a broad range of ideas, generate novel research hypotheses, a detailed research overview, and experimental protocols \cite{google_coscientist}, facilitating more robust ideation \cite{storey2024disruptive}. As demonstrated in a recent AI-augmented Brainwriting study \cite{shaer2024ai}, LLMs contribute to the divergence stage of ideation by introducing novel perspectives and generating diverse research angles that researchers might not have considered independently. This fosters creativity and mitigates common brainstorming barriers such as fixation and cognitive inertia.

Recent LLM-based contributions, such as the Disruptive Research Playbook~\cite{storey2024disruptive}, which helps to formulate socially relevant research questions to challenge assumptions and refine focus, and the AI Co-Scientist~\cite{google_coscientist}, which supports generating, debating, and refining hypotheses through a multi-agent system that allows iteratively improving research directions, point toward a future where LLM-based solutions evolve research goal and question formulation to accelerate impactful scientific discovery.

In a broader perspective, LLMs can also enable researchers to define more ambitious research goals that call for mixed-method research \cite{storey_guiding_2025}. LLMs can assist in parts of the research process where the researcher lacks experience or resources or when the total research design becomes complex. E.g., LLMs may advise the researcher in avoiding known anti-patterns in mixed method research such as `Sample contamination' and `Integration failure'~\cite{storey_guiding_2025}.

\textit{\textbf{Obsolece:}} LLMs can \textsc{reduce the manual effort in systematic literature reviews (SLRs)}, automating several time-consuming tasks. Recent research found that ChatGPT can support study selection and also help improve search string formulation by suggesting synonyms and relevant terms for more effective Boolean queries~\cite{felizardo2024chatgpt}. LLMs are already being used in other domains, such as medicine, to help extract~\cite{khraisha2024can} and synthesize data~\cite{lam2023comparing}, helping researchers identify key concepts and themes in large volumes of literature. LLMs also support inclusion/exclusion decisions, reducing manual effort while maintaining consistency. Human oversight remains critical despite these advantages, as LLMs can sometimes provide persuasive but inaccurate information. By using LLMs for preliminary automation and human validation of critical decisions, researchers can streamline the SLR process, reduce cognitive load, and focus on higher-level analysis while ensuring accuracy and reliability.

\textit{\textbf{Retrieve:}} LLMs can revive the culture of informal ideation and intellectual discourse in research, previously known as \textsc{``coffee house'' research} \cite{ellis2008introduction}. Through natural language interaction with LLMs and rapid information synthesis, these tools facilitate the rapid exploration and prototyping of research ideas before formalizing them into papers, which may take time to generate feedback. Furthermore, by reducing much of the tedious work involved in research, LLMs can allow researchers to reallocate that time to engage in more intellectual discourse, fostering deeper engagement in small discussion-based workshops. Potentially enhanced by AI-driven insights, these conversations can lead to a more thorough exploration of fundamental research questions, helping researchers refine their goals, develop innovative ideas, and uncover novel interdisciplinary connections. \footnote{It should be noted that historical coffee house research was not always inclusive. Women, for example, often did not attend~\cite{ellis2008introduction}. We envision a modern and more inclusive version of coffee house research being retrieved.} 

LLMs can also switch trending research topics. For example, since LLMs are used to automate more aspects of code generation and validation, we may see a resurgence of research interest in formal specification and verification research~\cite{meyer} to ensure that the outputs of LLMs are correct. We expect to also see an increased research focus on parts of the software development cycle where humans must remain in the loop. For example, requirements engineering and research to improve practices around understanding user needs may see a resurgence as code generation becomes more fully automated. Research can investigate how LLMs can help to manage natural language subjectivity to transform unstructured data into structured requirements, supporting automated elicitation, inconsistency detection, and traceability. 

\textit{\textbf{Reverse:}} While LLMs can accelerate various research tasks, their use in formulating research questions, defining study goals, and structuring investigations raises concerns about creativity stagnation. Ideally, research questions should be driven by intellectual curiosity, domain expertise, and an ability to challenge conventional wisdom—qualities that LLMs, by design, lack. Since LLMs today tend to generate content based on existing knowledge, relying too heavily on them for research ideation risks creating a \textsc{creativity echo chamber}, where generated questions and study designs reflect common patterns rather than genuinely novel insights (that is the opposite of the benefit of using LLMs for creative research directions). This effect is particularly concerning given that modern academic incentives often emphasize rapid publication over deeply impactful contributions. If LLMs reinforce established knowledge structures, researchers may unknowingly converge on ``safe'' and predictable topics, leading to a homogenization of research rather than groundbreaking discoveries.

\vspace{-2mm}
\subsection{Analysis \& Interpretation}

The analysis and interpretation of SE data present unique challenges when studying human and social aspects, requiring researchers to make sense of complex, qualitative, and often interrelated findings from multiple sources~\cite{bano2024large}. LLMs are now transforming how researchers approach this intricate analytical process. 

\textit{\textbf{Enhance:}} By processing vast amounts of data, LLMs can enhance \textsc{quantitative, qualitative, and mixed-methods analysis} by identifying trends, anomalies, correlations, and patterns that might be missed in manual analysis. 
In SE research, this applies to quantitative data (\textit{e.g.}, controlled study measurements, test results, defect reports, commit logs) and qualitative data (\textit{e.g.}, interviews, meeting transcripts). LLMs can assist in applying quantitative and qualitative research methods to analyze such data. In quantitative research, they can be used to analyze structured datasets to extract key patterns and relationships. In qualitative research, they have been applied to identify sentiment~\cite{10.1145/3697009}, themes\cite{mathis2024inductive,de2024performing}, and to support the application of a variety of qualitative research methods~\cite{BarrosAGKKNB25}. By enhancing efficiency and scalability, LLMs can reduce the time required for data analysis, alleviating manual effort in both quantitative and qualitative research. They can also improve consistency in calculations and coding while enhancing generalizability by enabling pattern identification across larger datasets and broader contexts. 
Furthermore, LLMs can facilitate mixed-methods research by integrating the visualization of qualitative and quantitative findings---a traditionally challenging task \cite{guetterman2015integrating}.

LLMs can integrate \textsc{diverse viewpoints} from a wide range of stakeholders –often overlooked in traditional research– by processing large volumes of data, including user feedback, developer discussions, and practitioner reports from forums, social media, and documentation repositories. This capability enables researchers to surface varied perspectives, identify emerging trends, and capture insights that span technical, social, and organizational contexts. By synthesizing this diverse input, LLMs can highlight patterns and conflicting opinions, and find emerging trends that enrich qualitative analysis with voices from end-users and practitioners.


\textit{\textbf{Obsolesce:}} LLMs are increasingly pushing \textsc{manual coding} aside in both qualitative and quantitative analyses. LLMs significantly reduce the time and effort required for manual annotation \cite{ahmed2024can}, which, when done with appropriate care to check semantic aspects and consistency, can improve efficiency in SE research. 

In qualitative studies, LLMs are already being used to code interview transcripts, documents, surveys, interviews, issue-tracker comments, and software reviews \cite{bano2024large}. In quantitative studies, LLMs can process large amounts of repository data to automatically classify code changes (e.g., bug fixes, refactorings, feature additions), extract and structure performance metrics, API usage patterns, and technical debt indicators from repositories, reducing the need for manual intervention. Additionally, by offering guidance on test selection, assumption checking, and result interpretation, LLMs can aid in \textsc{manually selecting and executing statistical techniques}—tasks that traditionally demanded significant expertise. 

\textit{\textbf{Retrieve:}} LLMs can revive interest in \textsc{finding related theories in other domains} to interpret SE findings (something we have not been doing enough of in recent years \cite{lorey2022social}).
By processing vast interdisciplinary literature, LLMs can foster information-seeking practices and facilitate analysis for interdisciplinary research \cite{zheng2024disciplink}, connecting SE challenges to established frameworks in fields such as cognitive psychology and organizational science, which could otherwise remain unexplored~\cite{wang_scientific_2023}. This retrieval of cross-domain knowledge extends beyond simple analogies, enabling researchers to recontextualize technical findings within broader theories, identify disciplinary intersections, and explore new research directions.

LLMs can bring back more \textsc{holistic, interdisciplinary studies} by uncovering historical patterns and forgotten theories from different domains to help SE researchers contextualize new qualitative and quantitative insights. By analyzing past literature, LLMs can potentially trace how SE theories and ideas developed over time and why some approaches became more popular~\cite{stol2024}. 
Tshitoyan et al.~\cite{tshitoyan2019nature} showed we can use AI to extract hidden patterns from scientific publications that predict future discoveries. Through this historical lens, researchers can gain a deeper understanding of ongoing challenges in the SE field and avoid reinventing the wheel.

\textit{\textbf{Reverse:}} Interdisciplinary research requires caution as over-reliance on LLMs without domain expertise can lead to misinterpretation and reduced research rigor. Over-reliance on LLMs can also cause \textsc{AI hallucinations}—fabricated and inaccurate output—leading to misleading interpretations.
Driven by probabilistic patterns, LLMs may generate plausible but incorrect conclusions, misattribute sources, or identify false patterns from noisy data, distorting findings and compromising research validity. One example is about using LLMs as annotators. While the model-to-model agreement can predict when LLMs could safely replace human annotators, it also highlights the potential for systemic bias where LLMs reinforce each other’s errors, creating a false perception of reliability~\cite{ahmed2024can}. Hence, while LLMs can support analysis and interpretation, human oversight is required. We need to keep in mind that LLMs currently cannot independently assess the validity of an argument, and critical thinking remains a human responsibility.
%






\vspace{-4mm}
\section{Call to Action}



Using LLMs for SE research presents a significant opportunity to speed up discovery and unlock new avenues for impactful research. Although caution is necessary to ensure that model biases, reliability, and ethical considerations, among other risks, are addressed, leaning into LLMs can streamline the research pipeline. Across all research stages (see cross-cutting impacts in Table~\ref{tab:tetrad_llms_se}), we see that using LLMs can \textsc{enhance research speed and creativity} and \textsc{reduce manual and tedious research tasks}. Thus, by strategically integrating LLMs, researchers can conduct more efficient, data-driven investigations, enabling faster insights into complex SE phenomena. However, \textsc{overreliance on LLMs can result in lower skills of researchers}. We must ensure that LLMs are used for research in a human-centric perspective to augment, and not replace, researchers. Balancing the accelerated pace enabled by LLMs with methodological rigor involving human oversight and critical thinking will ensure the research remains valid and transformative. 

We also believe that the accelerated pace of research enabled by LLMs presents a pivotal decision for our research community. Although LLMs can be used to accelerate paper production, their true transformative potential lies in enabling researchers to redirect their time toward deeper intellectual engagement and \textsc{focus on doing impactful research}. By freeing researchers from time-consuming tasks such as searching for the literature and initial data analysis, LLMs create space for more meaningful collaborative discussions about the impact of research and its social implications. 

We urge our fellow members of the software engineering research community to use any time saved by LLMs to invest in thoughtful dialogue with colleagues, stakeholders, and potential beneficiaries of their work. Such conversations can help identify pressing research questions and ensure that research addresses genuine societal needs rather than merely contributing to academic metrics. This shift from quantity to quality of research output could lead to more impactful and purposeful scientific contributions that better serve both the academic community and society at large.

The SE research community must take proactive steps to harness the LLMs' benefits while mitigating their risks. As these models become increasingly integrated into research workflows, structured guidelines, evaluation processes, and educational initiatives are essential to maintaining scientific rigor and reproducibility. Without clear guidance, there is a danger of overreliance, hidden biases, loss of the human's role in constructing theory, and compromised methodological integrity. To address these concerns, we propose four key actions: (1) experimenting with LLMs in SE research to build concrete experiences, (2) developing guidelines for transparent reporting and reviewing \cite{ralph2020acm},(3) establishing benchmarks for evaluating LLM-generated research artifacts, and (4) creating educational resources to train researchers in responsible LLM usage.

(1) We encourage curiosity and wide experimentation of LLMs in SE research, working with real cases and real data. We may see various effects on SE research as a practice across the research pipeline, but some of these are so far based on (qualified) assumptions, grounded on early and fragmented experience. We need to gain and share more experience to contrast and detail the issues we have started identifying and - most likely - identify new ones.

(2) We advocate for clear and transparent reporting on using LLMs in research, in combination with clear review guidelines about the use of LLMs in SE research.
Every study incorporating LLMs should explicitly document how these models influence study design, data collection, and analysis. This includes specifying the LLM model and version used, the exact prompting strategies employed (including any sensitivity analyses), and the mechanisms for human oversight. Without standardized reporting, it becomes hard to assess the validity of findings, compare results across studies, or detect systematic biases introduced by AI-generated outputs. Establishing clear disclosure standards will enhance the interpretability, reproducibility, and credibility of research involving LLMs \cite{wagner2025evaluationguidelinesempiricalstudies}.

(3) To address the high variability in LLM performance, we propose establishing and maturing benchmarks to evaluate the quality and reliability of LLM-generated research artifacts. This initiative should include a publicly available dataset covering a range of SE research tasks, such as annotation, summarization, and causal inference, to systematically assess where and how LLMs can be reliably used. Additionally, validation protocols should be established, leveraging metrics such as inter-rater agreement scores (e.g., Krippendorff's $\alpha$) and consistency checks across different prompts and models \cite{ahmed2024can}. By creating a standardized evaluation process, the research community can ensure that LLMs are deployed only in contexts where their reliability has been empirically demonstrated, reducing the risk of misleading or flawed research conclusions.



(4) We highlight the urgent need for educational resources to guide researchers in the responsible use of LLMs. Without proper training, early-career researchers risk losing foundational skills like critical thinking and hands-on analysis. To prevent this, we recommend instruction in prompt engineering, bias mitigation, and hybrid human-LLM workflows, supported by workshops and case studies. As the community begins to understand the challenges of integrating LLMs into SE research, efforts are underway to develop best practices and guidelines\footnote{Guidelines for using LLMs in SE Research - https://llm-guidelines.org/}.




By equipping researchers with LLM literacy, strong methodological foundations, structured reporting, robust validation frameworks, and targeted education, the SE research community can responsibly integrate LLMs as an aid rather than a substitute for critical inquiry, preserving essential research skills while safeguarding the rigor and reliability of empirical studies. While this is a position paper, 
future work could build on our discussion by grounding it in specific SE areas—such as program analysis or repair and software traceability—and how LLMs can change the review process, a concern across the entire community.

\vspace{-2mm}
\section{Acknowledgments} We thank Thomas Zimmermann for his essential contributions to this study. Our sincere appreciation also goes to the Alfred P. Sloan Foundation and the Carlsberg Foundation for the 2nd Copenhagen Symposium on Human-Centered SE and AI (G-2024-22586 and CF24-0693 to Daniel Russo), held at Aalborg University on Nov2024, the National Science and Research Council of Canada (NSERC) RGPIN-2025-6813, the Rutherford Discovery Fellowship administered by the Royal Society Te Apārangi, the DARE project (PNC0000002, CUP B53C22006420001), and the QualAI project (2022B3BP5S, CUP H53D23003510006).
\vspace{-2mm}

\bibliographystyle{ACM-Reference-Format}
\bibliography{references}


\end{document}